\documentclass[11pt]{article}
\pdfoutput=1

\usepackage{amstext}
\usepackage{amsfonts}
\usepackage{amssymb}

\usepackage[pdftex]{hyperref}

\def\ben{\begin{equation}}
\def\een{\end{equation}}
\def\half{{\textstyle{1\over2}}}

\let\a=\alpha \let\b=\beta \let\g=\gamma \let\d=\delta \let\e=\varepsilon
  \let\q=\theta \let\k=\kappa
     \let\r=\rho
\let\s=\sigma \let\t=\tau

\let\w=\omega \let\G=\Gamma

\let\pa=\partial
\def\be{\begin{equation}}
\def\ee{\end{equation}}
\def\beq{\begin{equation}}
\def\eeq{\end{equation}}
\def\ba{\begin{array}}
\def\ea{\end{array}}

\def\dalemb#1#2{{\vbox{\hrule height .#2pt
       \hbox{\vrule width.#2pt height#1pt \kern#1pt
               \vrule width.#2pt}
       \hrule height.#2pt}}}

\newcommand{\bea}{\begin{eqnarray}}
\newcommand{\eea}{\end{eqnarray}}

\def\R{{{\Bbb R}}}

\def\Lag{{\mathcal{L}}}

\begin{document}
\begin{center}

{ \LARGE {\bf Spectral weight in holographic scaling geometries}}

\vspace{1cm}

{\large Sean A. Hartnoll and Edgar Shaghoulian}

\vspace{0.7cm}

{\it Department of Physics, Stanford University, \\
Stanford, CA 94305-4060, USA \\}

\vspace{1.6cm}

\end{center}

\begin{abstract}

We compute the low energy spectral density of transverse currents in theories with holographic duals that exhibit an emergent scaling symmetry characterized by dynamical critical exponent $z$ and hyperscaling violation exponent $\theta$. For any finite $z$ and $\theta$, the low energy spectral density is exponentially small at nonzero momentum. This
indicates that any nonzero momentum low energy excitations of putative hidden Fermi surfaces are not visible in the classical bulk limit. We furthermore show that if the limit $z \to \infty$ is taken with the ratio $\eta = - \theta/z > 0$ held fixed, then the resulting theory is locally quantum critical with an entropy density that vanishes at low temperatures as $s \sim T^\eta$. In these cases the low energy spectral weight at nonzero momentum is not exponentially suppressed, possibly indicating a more fermionic nature of these theories.

\end{abstract}

\pagebreak
\setcounter{page}{1}

\section{Introduction}

\subsection{Context}

The notion of a Fermi surface is a conceptual cornerstone of condensed matter physics.
At a first glance, this might appear surprising because the simplest picture one has of a Fermi surface
is intrinsically perturbative: the momentum space description of a density of non-interacting particles obeying Pauli exclusion. It turned out, however, that Fermi surfaces are rather robust. Firstly, they are self-consistent stable IR fixed points of renormalization group flow (up to BCS-type instabilities) in the absence of critical bosonic degrees of freedom \cite{Polchinski:1992ed, Shankar:1993pf}. Secondly, the existence of singularities at nonzero momenta in certain fermionic correlators can be nonperturbatively inferred from the Luttinger theorem; see \cite{Huijse:2011hp} for a recent discussion.

Many of the most interesting cases, however, are precisely characterized by the presence of gapless bosonic excitations that drive a renormalization group flow to a strongly coupled IR regime, e.g. \cite{sungsik, Metlitski:2010pd, Metlitski:2010vm}. The low energy theory will typically not have well-defined fermionic quasiparticles. Furthermore, if the IR fermions are charged under an emergent gauge symmetry, it may not be clear, from an IR perspective, which gauge-invariant fermionic correlators one is supposed to look at in order to detect momentum space singularities \cite{Huijse:2011hp}. Clean examples of this situation have emerged in holographic descriptions of finite density matter: when all the bulk electric flux emerges from behind a horizon, gauge-invariant fermionic probes do not have singularities  at finite momentum, see \cite{Hartnoll:2011fn} and references therein. The momentum space singularities found in \cite{Lee:2008xf, Liu:2009dm, Cubrovic:2009ye, Faulkner:2009wj} are associated with a fraction of the flux that is sourced by charge outside the horizon \cite{Iqbal:2011bf}.

An important question is whether even in these `fully fractionalized' \cite{Hartnoll:2011pp} cases, without obvious singularities in fermion correlators, one can associate `fermionic' characteristics to the finite density matter. An interesting property of systems with a Fermi surface is a logarithmic violation of the area law of the entanglement entropy \cite{one,two,three,four,five}. It was recently found that a specific class of holographic geometries with flux emanating from behind a (in general singular\footnote{Remarkably, the limiting case with $\theta=1$ and $z=3/2$ (in two space dimensions) has a regular geometry in the IR \cite{Shaghoulian:2011aa}. It is tempting to connect this observation with the fact that this is the value of $z$ appearing in some field theoretic treatments \cite{Metlitski:2010pd}.}) horizon exhibited such logarithmic area law violation \cite{Ogawa:2011bz}.  It was subsequently shown that these same geometries have further characteristics reminiscent of Fermi surfaces, including the fact that the dependence of the entropy upon temperature indicates that the critical excitations are effectively one dimensional \cite{Huijse:2011ef}. The `scaling geometries' in question will be reviewed shortly, but are characterized by a dynamical critical exponent $z$ and a hyperscaling violation exponent $\theta$. The extremely interesting claim of \cite{Ogawa:2011bz, Huijse:2011ef} is that geometries with $\theta = d-1$ (throughout this paper we will restrict to two spatial dimensions, so $d=2$) dually describe fractionalized Fermi surfaces. Further aspects of these spacetimes were studied in \cite{Dong:2012se}.

\subsection{Results}

A compelling aspect of the papers \cite{Ogawa:2011bz, Huijse:2011ef} is that the Fermi surface-like physics is visible at leading order in the bulk semiclassical limit. One objective of this paper will be to highlight a feature of these classical scaling geometries that complicates their interpretation as describing hidden Fermi surfaces. A defining feature of Fermi surfaces is that the low energy fermionic excitations do not live at the origin of momentum space but rather at finite momenta. While, as discussed above, we are interested in cases where we cannot probe these fermions directly, the existence of low-lying degrees of freedom at finite momentum can be detected indirectly, for instance by looking at current-current or density-density correlators. Here the currents and densities could refer either to the electric charge or to the energy. Most dramatically, we might hope to see `$2 k_F$' singularities, assuming these can survive strong interactions, but in fact there is a more basic issue.

On the face of it, the scaling symmetry of the geometries is $t \to \lambda^z t$, $x \to \lambda x$, and so, for any finite $z$, IR degrees of freedom are scaled towards the origin in momentum space. Therefore, we can expect that at low energies, any finite momentum current or density excitations will be very off shell and consequently exponentially suppressed. The first result in this paper will be to show this by an explicit computation. For technical reasons, we are only able to obtain (\ref{eq:result}) explicitly in cases where $z+\theta=4$, where the equations simplify slightly. We will compute the spectral density of the transverse current operator and obtain
\be\label{eq:result}
\text{Im} \, G^R_{J_\perp J_\perp}(\w,k) \; \sim \; \exp\left\{- a^2 \left(\frac{k^z}{\w}\right)^{1/(z-1)} \right\} \,, \qquad (\w \to 0, k \text{ fixed}) \,.
\ee
Here $a^2$ is a computable positive number. Rather similar results have been previously obtained for fermionic spectral densities \cite{Faulkner:2010tq, Iizuka:2011hg, Hartnoll:2011dm, Iqbal:2011in}.
The above expression holds true for any $\theta$ at finite $z < \infty$ and does not involve any limit other than low frequency. In particular it is true for $\theta = 1$. The upshot is that there are essentially no low energy charged degrees of freedom at finite momentum. The restriction to charged degrees of freedom is not crucial, as the same result holds for e.g. the energy current. The spectral densities of these spacetimes look more bosonic than fermionic. It is clearly an important question to understand how this statement connects with the nontrivial entanglement entropy uncovered by  \cite{Ogawa:2011bz, Huijse:2011ef}.

The emphasis on spectral density at low energy but finite momentum is more than a theoretical quibble.
A central property of Fermi liquids is that their DC resistivity in the presence of umklapp scattering goes like
$T^2$ at low temperatures. Efficient (i.e. power law at low temperature) umklapp scattering is only possible if the system has a significant low energy spectral weight at finite momentum, otherwise there are too few modes for the lattice to scatter \cite{Hartnoll:2012rj}. Results such as (\ref{eq:result}) are connected to the fact that in the presence of umklapp scattering, holographic theories with scaling geometry duals will have a DC resistivity that is exponentially small in temperature, i.e. Boltzman suppressed to go like $e^{- k_L^z/T}$, with $k_L$ the lattice momentum. This is a physical sense in which such systems behave differently to Fermi surfaces.

If we remain within the framework of the scaling geometries considered thus far in the literature,
the above conclusions can only be invalidated by taking the limit $z \to \infty$. In this limit, position does not scale and so there are low energy modes at all momenta. Indeed we see that the exponent in (\ref{eq:result}) is not parametrically large in this limit. The entropy density at finite temperature scales like $s \sim T^{(2-\theta)/z}$ \cite{Hartnoll:2011fn, Huijse:2011ef} and so if the $z \to \infty$ limit is taken with any finite $\theta$, then the resulting phase has a nonzero ground state entropy density. This fact has been widely discussed in the context of extremal black holes and may be undesirable. The second result in this paper is to note that there is an interesting double scaling limit which leads to locally critical theories with $z=\infty$, but which have a vanishing zero temperature entropy density. Namely, take
\be\label{eq:fixed}
z \to \infty \quad \text{with} \quad  \frac{-\theta}{z} \equiv \eta > 0 \quad \text{fixed} \,.
\ee
All physical quantities behave nicely in this limit.
In particular, the spacetime geometry is conformally related to $AdS_2 \times \R^2$ and the entropy density goes like $s \sim T^{\eta}$. We will compute the transverse current spectral density in this limit and show that it is not exponentially suppressed at low energies and finite momentum. Rather, we find
\be
\text{Im} \, G^R_{J_\perp J_\perp}(\w,k) \; \propto \; \w^{2 \nu_-(k)} \,,\qquad (\w \to 0, k \text{ fixed}) \,,
\ee
with the exponent $\nu_-(k)$ dependent on $k$ and $\eta$ as we will determine below.
Consequently the DC resistivities due to umklapp scattering can be expected to be power law in temperature \cite{Hartnoll:2012rj}. In this sense, this class of theories (including $AdS_2 \times \R^2$) might be considered the most fermionic of known gravity duals.

\section{Theory and backgrounds}

In order to perform concrete computations, we need a bulk theory that has the scaling geometries as solutions.
A convenient model to consider is the following Einstein-Maxwell theory with a dilatonic scalar field
\be\label{eq:lag}
\Lag = \frac{1}{2\k^2} R - \frac{1}{4 e^2} Z(\Phi) F^2 - \frac{1}{\k^2} \left( \pa \Phi \right)^2 - \frac{1}{2 \k^2 L^2} V(\Phi) \,.
\ee
We will take the two functions of the dilaton to be
\be
Z(\Phi) = Z_0^2 e^{\a \Phi} \,, \qquad V(\Phi) = - V_0^2 e^{- \b \Phi} \,,
\ee
where instead of $\a$ and $\b$ it will shortly become helpful, following \cite{Huijse:2011ef}, to frame our discussion in terms of $z$ and $\theta$ defined by
\be
\q = \frac{4 \b}{\a + \b} \,, \qquad z = \frac{16 + \a^2 + 2 \a \b - 3 \b^2}{\a^2 - \b^2} \,.
\ee
This class of bulk theories has been used to model aspects of QCD at finite density \cite{DeWolfe:2010he} as well as quantum critical condensed matter systems \cite{Taylor:2008tg, Goldstein:2009cv, Charmousis:2010zz, Cadoni:2011kv}. The interesting fact from the quantum critical perspective is that these theories admit scaling solutions. Write the metric and Maxwell field as
\be\label{eq:metric}
ds^2 = L^2 \left(- f(r) dt^2 + g(r) dr^2 + \frac{dx^2 + dy^2}{r^2} \right) \,, \qquad A = \frac{eL}{\k} h(r) dt \,.
\ee
Then one finds solutions to the equations of motion following from (\ref{eq:lag}) of the form \cite{Huijse:2011ef}
\bea\label{eq:scaling}
f = r^{- 2 (\q-2z)/(\q-2)} \,, & \quad g = g_0 \, r^{- 4 (\q -1)/(\q - 2)} \,, \\
h = h_0 \, r^{2(2+z-\q)/(\q-2)} \,, & \quad \Phi = {\textstyle \sqrt{\frac{2z-2-\q}{2-\q}} } \log r \,, \nonumber
\eea
where the constants are
\be
g_0 = \frac{4 (1+z-\q)(2+z-\q)}{V_0^2(2-\q)^2} \,, \qquad h_0^2 = \frac{z-1}{Z_0^2 (2+z-\q)} \,.
\ee
In these coordinates $r \to 0$ is the near-boundary UV while $r \to \infty$ is the IR interior.

It was emphasized in \cite{Gouteraux:2011ce, Huijse:2011ef} that scale transformations act on the scaling solution as
\bea
\vec x & \to & \lambda \vec x \,,  \\
t & \to & \lambda^z t \,, \\
ds & \to & \lambda^{\theta/2} ds \,, \\
r & \to & \lambda^{(2-\theta)/2} r \,. 
\eea
Thus $z$ may be interpreted as the dynamical critical exponent while $\theta$ is the `hyperscaling violation exponent'. In addition, in these models, there is a specific logarithmic violation of scaling by the dilaton and the Maxwell potential $A \to \lambda^{\theta-2}A$. As we noted in the introduction, the Fourier transform of these scalings implies that, if $z>1$ is finite, low energies corresponds to low momenta.

The scaling geometries (\ref{eq:scaling}) may be considered as the IR limit of an asymptotically $AdS_4$ spacetime held at a finite chemical potential, or as solutions in their own right. All of the conserved electric flux, $\int_{\R^2} \star (e^{\a \Phi}  F)$, emanates from behind the `horizon' at $r \to \infty$ and therefore these geometries are `fully fractionalized' in the sense of \cite{Hartnoll:2011pp}. Any charged bulk fermions (or scalars) are assumed to be sufficiently heavy as to not produce a charge density.

\section{Transverse current perturbations}

We would like to compute a spectral density that probes the existence of charged excitations, and in fact also neutral excitations, at low energy $\w$ but finite momentum $k$. Perturbing the scaling solution described above at finite momentum and energy leads to a complicated set of coupled equations. The simplest set of equations that can be consistently decoupled are those describing the transverse channel. If we take the momentum of the perturbations to be in the $x$ direction, then in the transverse channel the following fields are excited: $\{\d A_y, \d g_{yt}, \d g_{xy}\}$. These modes are dual to excitations of transverse electric and heat currents in the field theory. The modes decouple from the remainder because they are odd under $y \to -y$.

\subsection{Gauge-invariant equations}

The transverse modes can be grouped into two gauge-invariant combinations, following for example \cite{Edalati:2010hk},
\be
\psi_1 = \d A_y \,, \qquad \psi_2 = r^2 \left(\w \, \d g_{xy} + k \, \d g_{yt} \right) \,.
\ee
Here we are taking all modes to be of the form
\be
\d X(r,t,x) = \d X(r) e^{- i \w t + i k x} \,.
\ee
The perturbed Einstein-Maxwell-dilaton equations following from the action (\ref{eq:lag}) about the scaling background (\ref{eq:scaling}) can be written as two coupled equations for $\psi_1$ and $\psi_2$. To make the equations more transparent, it is useful to perform several rescalings. Firstly, rescale the gauge invariant fields
\be
\psi_1(r) = \frac{(z-1)}{Z_0^2 h_0}  \frac{1}{\w} \left(\frac{k}{\w}\right)^{(\theta-3)/(z-1)}  \phi_1(r) \,, \qquad \psi_2(r) = \phi_2(r) \,.
\ee
Secondly, collect the momentum dependence into the following quantity
\be\label{eq:kap}
\kappa = \sqrt{g_0} (2-\theta) \left(\frac{k^z}{\w} \right)^{1/(z-1)} \,,
\ee
and redefine the radial variable to be
\be
r = \left(\frac{\w}{k}\right)^{(\theta-2)/2(z-1)} \rho^{2-\theta} \,.
\ee
Note that we are assuming that $\theta \leq 1$ \cite{Huijse:2011ef}. The perturbation equations for $\phi_1$ and $\phi_2$ as a function of $\rho$ are then written as
\bea
\lefteqn{\phi_1'' - \frac{2z+2\theta-9}{\r} \phi_1' + \frac{2 \r^{2z+2\theta-5}}{\r^{4z} - \r^4} \phi_2' } \nonumber \\
&& \qquad + \left(\k^2 (\r^{4z} - \r^4) - \frac{8 (2+z-\theta)(z-1) \r^{4z}}{\r^{4z} - \r^4} \right) \frac{1}{\r^2} \phi_1 = 0 \,, \label{eq:p1}
\eea
and
\bea
\lefteqn{\phi_2'' - \frac{(3+2z-2\theta) \r^{4z} + (2z+2\theta-7)\r^{4}}{\r (\r^{4z} - \r^4)} \phi_2' - 4 (z-1)(2+z-\theta) \r^{7-2z-2\theta} \phi_1' }\nonumber \\
&& \qquad + \, \k^2 \frac{\r^{4z} - \r^4}{\r^2} \phi_2 + (z-1)^2 (2+z-\theta)\frac{16 \r^{2z-2\theta+6}}{\r^{4z} - \r^4} \phi_1 = 0 \,. \label{eq:p2}
\eea
These equations follow from the Einstein-Maxwell-dilaton equations for $\{\d A_y, \d g_{yt}, \d g_{xy}\}$.

In the case of pure Einstein-Maxwell theory, it is possible to decouple these equations and obtain two decoupled second order equations \cite{Kodama:2003kk,Edalati:2010hk}. We have not been able to find an analogous decoupling for the Einstein-Maxwell-dilaton case in general. In a later section, we will decouple the equations in the double scaling limit mentioned in the introduction.

It will also be useful below to observe that the equations are associated with the following conserved flux
\be\label{eq:flux}
j = \frac{\r^{2z + 2\theta-3}}{\g^2 \left(\r^{4z} - \r^4\right)} \left(\overline \phi_2 \phi'_2
- \phi_2 \overline \phi'_2 \right) + \r^{9 - 2z-2\theta} \left(\overline \phi_1 \phi'_1
- \phi_1 \overline \phi'_1 \right) + \frac{2 \rho^4}{\rho^{4z} - \rho^4} \left(\overline \phi_1 \phi_2 - \phi_1 \overline \phi_2 \right) \,,
\ee
where $\g^2 = 2(z-1)(2+z-\theta)$. This flux satisfies $dj/d\rho = 0$.

\subsection{Solving the equations at low frequency and fixed momentum}

In equations (\ref{eq:p1}) and (\ref{eq:p2}) we see that all of the energy and momentum dependence has been scaled into the single parameter $\kappa$, defined in (\ref{eq:kap}). The remaining terms in the equation depend only on the constants $z$ and $\theta$. From this observation, it is immediate that to leading order as $\w \to 0$, the spectral density cannot have nonanalytic structure at finite momentum (which would have been reminiscent of $2k_F$ singularities). This is because the $k$ dependence will be tied, through $\kappa$, to the leading order $\w$ dependence.

At the low energies and fixed momentum of interest,
\be
\kappa \gg 1\qquad (\w \to 0,\, k \text{ fixed})\,.
\ee
In this limit, the parameter $\k$ acts precisely as a WKB parameter. Away from turning points we can look for solutions of the form
\be\label{eq:wkb}
\phi_i(\r) = a_i(\r) \, e^{\k S(\r)} \,. 
\ee
To leading order at large $\k$, the equations of motion give
\be
S(\r) = \pm i  \int \frac{d\r}{\r} \sqrt{\r^{4z} - \r^4} \,.
\ee
It is clear that the WKB expression will break down close to $\r \approx 1$. This is not quite a standard turning point, however, as the equations (\ref{eq:p1}) and (\ref{eq:p2}) contain additional terms that are singular at $\r \approx 1$. We also see that the WKB solution will be problematic near the boundary $\r \approx 0$.

To obtain the retarded Green's functions of the field theory operators dual to $\phi_1$ and $\phi_2$, and hence the desired spectral densities, we must obtain the near boundary behavior of the fields given infalling boundary conditions at the horizon \cite{Hartnoll:2009sz}. We proceed to do this via a sequence of matchings between regimes in which the WKB form (\ref{eq:wkb}) is valid.

In the near horizon IR regime, with $\r > 1$, the WKB solution is oscillating (we assume that $z>1$), as we should expect. There are two linearly independent solutions satisfying infalling boundary conditions at the horizon
\be\label{eq:in}
\phi_i = a_i(\r) \exp\left\{ i \k \int_1^{\r} \frac{d\s}{\s} \sqrt{\s^{4z} - \s^4} \right\} \,, \qquad (\r > 1)\,.
\ee
The two free constants are contained in the functions $a_i(\r)$.
We now need to match this solution onto the solution in the exponential regime away from the horizon at $\r < 1$. In this region the solution takes the form:
\be
\phi_i = b^+_i(\r) \exp\left\{\k \int_{\r}^1 \frac{d\s}{\s} \sqrt{\s^{4} - \s^{4z}} \right\}
+ b^-_i(\r) \exp\left\{- \k \int_{\r}^1 \frac{d\s}{\s} \sqrt{\s^{4} - \s^{4z}} \right\} \,, \qquad (\r < 1) \,. \label{eq:bnd}
\ee
We will give an explicit form for the $b_i^{\pm}(\r)$ shortly. In general these would depend on a total of four free constants; we now proceed to fix two of these by matching onto the infalling solution (\ref{eq:in}).

We can match the WKB solution onto a solution valid in the matching region $\r \approx 1$ by taking the following scaling limit of the perturbation equations. Let
\be
\t = \k^{2/3} (\r - 1) \,,
\ee
and take $\k \to \infty$ keeping $\t$ fixed. This clearly zooms us into the $\r \approx 1$ region. Large $|\t|$ will take us back into the WKB regions while $|\t|$ small, or order one, is outside of the WKB regime of validity. The power $\k^{2/3}$ is chosen to balance the $\k^2$ terms in the perturbation equations with the second derivative terms. In this scaling limit, the equations (\ref{eq:p1}) and (\ref{eq:p2}) become
\bea
\ddot \phi_1 + \frac{1}{2 (z-1) \t} \dot \phi_2 + 4 (z-1) \t \phi_1 & = & 0 \,, \label{eq:aa} \\
\ddot \phi_2 - \frac{1}{\t} \dot \phi_2 + 4 (z-1) \t \phi_2 & = & 0 \,. \label{eq:bb}
\eea
Dots denote derivatives with respect to $\t$.
These equations can be solved in terms of Airy functions and derivatives of Airy functions. This is rather similar to the standard WKB matching procedure. The general solution to equations (\ref{eq:aa}) and (\ref{eq:bb}) is
\bea
\phi_1 & = & c_1 \, \text{Ai}(-2^{2/3} (z-1)^{1/3} \, \t) + c_2 \, \text{Bi}(-2^{2/3} (z-1)^{1/3} \, \t) + \frac{1}{2(1-z)} \phi_2 \,, \label{eq:airy0} \\
\phi_2 & = & c_3 \, \text{Ai}'(- 2^{2/3} (z-1)^{1/3} \, \t ) + c_4\, \text{Bi}'(- 2^{2/3} (z-1)^{1/3} \, \t ) \,. \label{eq:airy}
\eea
Expanding these solutions for $\t \to + \infty$ and matching onto the infalling solutions (\ref{eq:in}) gives
\be\label{eq:c1c2}
c_3 = i c_4 \,, \qquad c_1 = i c_2 \,. 
\ee
In this matching, recall that $z>1$, so there are no phases in the arguments of the Airy functions.
We can in principle further obtain $\{c_2,c_4\}$ in terms of the constants characterizing the near horizon solution, but there is no advantage to doing so. The above relations ensure that the scaling solution becomes infalling. One can check that the scaling solution solves the full equations of motion at $\t \sim 0$, to leading order at large $\k$, and is therefore suitable for interpolation through to $\t \to - \infty$.

Expanding the Airy functions (\ref{eq:airy0}) and (\ref{eq:airy}) as $\t \to - \infty$ with the constants related by (\ref{eq:c1c2}), one finds that the infalling near horizon solutions map onto a sum of exponentially growing and decaying solutions. Specifically, the solution takes the WKB form (\ref{eq:bnd}) with the functions $b_i^{\pm}(\r)$ given by \pagebreak
\bea\label{eq:bmzero}
b^+_1 & = & \frac{1}{\g} \frac{\r^{z+\theta-4}}{\left(\r^4 - \r^{4z} \right)^{1/4}}
\left(\hat c_2 \cos x(\r) + i \, \hat c_4 \sin x(\r)   \right) \,, \\
b^-_1 & = &  \frac{i}{2} \frac{1}{\g} \frac{\r^{z+\theta-4}}{\left(\r^4 - \r^{4z} \right)^{1/4}}
\left( \hat c_2 \cos x(\r) - i\, \hat c_4 \sin x(\r)   \right) \,, \\
b^+_2 & = & \r^{2-z-\theta} \left(\r^4 - \r^{4z} \right)^{1/4} \left(i \, \hat c_2 \sin x(\r) + \hat c_4 \cos x(\r)   \right)  \,, \\
b^-_2 & = & \frac{i}{2} \r^{2-z-\theta} \left(\r^4 - \r^{4z} \right)^{1/4}  \left(i \, \hat c_2 \sin x(\r) - \hat c_4  \cos x(\r)   \right)  \,,
\eea
where as before $\g = \sqrt{2 (z-1) (2+z-\theta)}$, and
\be
x(\rho) =\sqrt{\frac{2+z-\q}{2(z-1)}} \arccos \r^{2-2z} \,.
\ee
The constants in (\ref{eq:c1c2}) are
\be\label{eq:ccc}
\{\hat c_2, \hat c_4\} \; \propto \; \{c_2, \k^{1/3} c_4\} \,.
\ee
The constant of proportionality here is not necessary to obtain the Green's functions we are after. We will shortly find that $\hat c_2 \sim \hat c_4$, so the rescaling by $\k^{1/3}$ in (\ref{eq:ccc}) indicates that $\phi_2 \ll \phi_1$ in the scaling regime with solutions (\ref{eq:airy0}) -- (\ref{eq:airy}). This implies that in fact equations (\ref{eq:aa}) and (\ref{eq:bb}) decouple. To find the explicit forms for the above functions $b^\pm_i$, it was necessary to solve the perturbation equations at subleading order in the WKB expansion ($\k \to \infty$). This required some educated guesswork, using the fact that the flux (\ref{eq:flux}) should be independent of $\r$.

Finally, we must match onto a solution that is valid near the asymptotic boundary, where WKB again breaks down.
Similar to the matching across the turning point, we take a scaling limit of the perturbation equations. In this case let
\be
\hat \t = \k^{1/2} \r \,.
\ee
To get all the way to the boundary in this scaling regime, inspection of the perturbation equations shows that one must also rescale one of the fields
\be
\phi_3 = \k^{4-z-\theta} \phi_2 \,.
\ee
One then takes $\k \to \infty$ with $\hat \t$ and $\phi_3$ held fixed. The following equations are obtained
\bea
\ddot \phi_1 + \frac{9-2z-2\theta}{\hat \t} \dot \phi_1 - \hat \t^2 \phi_1 & = & 2 \hat \t^{-9+2z+2\theta} \dot \phi_3 \,, \label{eq:b1} \\
\ddot \phi_3 - \frac{7-2z-2\theta}{\hat \t} \dot \phi_3 - \hat \t^2 \phi_3 & = & 4(z-1)(2+z-\theta) \hat \t^{7-2z-2\theta} \dot \phi_1 \label{eq:b2} \,.
\eea
Here dot indicates differentiation with respect to $\hat \t$. These equations match onto the WKB regime at large $\hat \t$ and have the same near boundary behavior as the full equations. Namely, as $\hat \t \to 0$, $\phi_1 \sim \hat \t^a$ with $a = \{0, 4(z-1),2(z+\theta-2), -2(2+z-\theta)\}$ while $\phi_3 \sim \hat \t^b$ with $b=\{0,2(2+z-\theta),-4(z-1),-2(z+\theta-6)\}$.

We have not been able to solve the near-boundary scaling perturbation equations (\ref{eq:b1}) -- (\ref{eq:b2}) in general. One can obtain equations in Schr\"odinger form, i.e. with no first derivatives, by the transformation
\bea
\phi_1 & = & \hat \t^{-9/2+z+\theta} \left( \hat \t^{\g} \phi_+ + \hat \t^{-\g} \phi_- \right) \,, \\
\phi_3 & = & \g \, \hat \t^{7/2-z-\theta} \left( \hat \t^{\g} \phi_+ - \hat \t^{-\g} \phi_- \right) \,.
\eea
Here, as above, $\g = \sqrt{2 (z-1) (2+z-\theta)}$.
In general the equations remain coupled and difficult to solve. Exceptional cases arise, however, whenever
\be\label{eq:restrict}
z + \theta = 4 \,.
\ee
In the remainder of this section we will consider only these cases. These include the case $z=3$, $\theta = 1$ which will have a logarithmically scaling entanglement entropy as found in \cite{Ogawa:2011bz}.
In the cases satisfying (\ref{eq:restrict}), the two components of the near boundary scaling Schr\"odinger equation decouple and become
\be
\ddot \phi_\pm = \hat \t^2 \phi_\pm + \frac{(4 \theta -11)(4 \theta - 13)}{4 \hat \t^2} \phi_\pm \,.
\ee
These equations are solved by the modified Bessel functions
\be\label{eq:bsol}
\phi_\pm = \pi \, d^\pm \hat \t^{1/2} I_{3-\theta}\left(\frac{\hat \t^2}{2} \right) + e^\pm \hat \t^{1/2} K_{3-\theta}\left(\frac{\hat \t^2}{2} \right) \,.
\ee
By expanding these Bessel functions as $\hat \t \to \infty$ we can match onto the WKB solutions given by (\ref{eq:bnd}) and (\ref{eq:bmzero}). We find that the constants are
\bea
d^+ & \propto & i e^{i \pi \theta/2} \k^{\theta-8/3} e^{-\k X} \left(\hat c_4 - \hat c_2 \right) \,, \\
d^- & \propto & - 4 i e^{i \pi \theta/2} \k^{10/3-\theta} e^{-\k X} \left(\hat c_4 + \hat c_2 \right)  \,, \\
e^+ & \propto & - \k^{\theta-8/3} \left( e^{-i \pi \theta/2} e^{-\k X} \left(\hat c_4 - \hat c_2 \right) 
+ 2 e^{i \pi \theta/2} e^{\k X} \left(\hat c_4 + \hat c_2 \right)\right) \,, \\
e^- & \propto & 4 \k^{10/3 - \theta} \left( e^{-i \pi \theta/2} e^{-\k X} \left(\hat c_4 + \hat c_2 \right) 
+ 2 e^{i \pi \theta/2} e^{\k X} \left(\hat c_4 - \hat c_2 \right)\right) \,.
\eea
Here
\be
X = \int_0^1 \frac{d\s}{\s} \sqrt{\s^4 - \s^{4z}} = \frac{\sqrt{\pi}}{4z} \frac{\G\left(\frac{1}{2(z-1)} \right)}{\G\left(\frac{z}{2(z-1)} \right)} \,,
\ee
 and the constant of proportionality is the same for the four constants $\{d^\pm, e^\pm\}$. This same integral appeared first in \cite{Faulkner:2010tq}.

Finally, we are able to obtain the retarded Green's function. The solutions (\ref{eq:bsol}) expand near the boundary, $\hat \t \to 0$, as
\be
\phi_\pm = 2^{5 - 2\theta} \Gamma \left(3-\theta\right) e^\pm \hat \t^{2\theta - 11/2} + 2^{2\theta-7} \left(
\frac{2\pi}{\G(4-\theta)} d^\pm + \G(\theta-3) e^\pm \right) \hat \t^{13/2-2 \theta} + \cdots \,. \label{eq:bex}
\ee
Here we see that the exponentially growing and decaying WKB modes have nontrivially mapped onto a mix of power law growing and decaying modes near the boundary. This seems to indicate that we could not have skipped the last step of solving the near boundary scaling equations (\ref{eq:b1}) and (\ref{eq:b2}), that restricted us to the cases with $z+\theta=4$.
The retarded Green's function is read off from the near boundary expansion (\ref{eq:bex}) by comparing to
\be
\phi_\pm = r^{(2\theta - 11/2)/(2-\theta)} + {\mathcal{G}}^R_\pm(\w,k) r^{(13/2 - 2\theta)/(2-\theta)} \,.
\ee
We have reintroduced the original radial coordinate $r$.
In computing the ${\mathcal{G}}^R_\pm$ Green's function, we must set the source for the $\phi_\mp$ mode to zero. That is, $\hat c_2$ and $\hat c_4$ must be related such that $e^\mp = 0$:
\be
\hat c_4 = \pm \hat c_2 (1 - e^{- i \pi \theta} e^{- 2 \k X} + \cdots) \,.
\ee
It is necessary to keep the subleading term here. The result for the spectral density is, to leading exponential order in the limit $\k \to \infty$,
\be\label{eq:final}
\text{Im} \, {\mathcal{G}}^R_\pm  \quad \propto  \quad k^{6 - 2 \theta} \, \exp\left\{- a^2 \left(\frac{k^z}{\w}\right)^{1/(z-1)} \right\} \,. \\
\ee
Here $a^2 = 4 \sqrt{g_0} (2-\theta) X$. Note that there is an extra factor of 2 in the exponent relative to what one might have guessed from the usual WKB tunneling amplitude. That is to say, the suppression due to the need to tunnel into the near horizon region is $e^{- 4 \k X}$. We have not been overly careful about the overall normalization of (\ref{eq:final}) -- while we have solved the equations in enough detail to get the coefficient right, a proper treatment requires a holographic renormalization approach.

Equation (\ref{eq:final}) is the advertised exponential suppression of the spectral weight. The Green's function obtained is that of the operators dual to the modes $\phi_\pm$ in the scaling geometry, that have a well defined scaling near the boundary of this geometry. If the scaling geometry describes the IR of, for instance, an asymptotically AdS spacetime at finite charge density, then these spectral densities will control the low frequency limit of the transverse electric and heat current spectral densities of the full theory, as in e.g. \cite{Edalati:2010hk}. In particular, the spectral densities of the currents will inherit the exponential suppression. It is possible that the low frequency matching of the IR Green's function to the full answer could introduce a singularity at finite momentum. For our bosonic fields, however, this would almost certainly indicate the presence of an instability at low momenta, a topic we will return to in the discussion, and in any case would be a property of the spacetime away from the scaling regime and therefore independent of the values of $\theta$ and $z$.

We have not found the Green's function for arbitrary finite $z$ and $\theta$; the obstacle is solving the equations (\ref{eq:b1}) and (\ref{eq:b2}) that match the WKB regime to the boundary. It seems clear, however, that the physical effect underlying the exponential suppression of spectral weight at nonzero momentum and low frequencies, namely the fact that the scaling symmetry is towards the origin in momentum space, applies equally to the general case. The cases we have solved, with $z + \theta = 4$, provide an explicit confirmation of this intuition. It is of course of interest to confirm the remaining cases explicitly also, if necessary numerically, in case some subtlety arises.

\section{Local criticality without a ground state entropy density}

In order to obtain a spectral density that is not exponentially suppressed at low energies and finite momenta, we now consider the `locally critical' limit $z \to \infty$. In order to avoid landing with a ground state entropy density, we will simultaneously take the limit $\theta \to - \infty$ with $\theta/z \equiv - \eta$ fixed. This double scaling is not as contrived as it may appear; keeping $\theta/z$ rather than $\theta$ fixed can be thought of as keeping a momentum rather than an energy fixed in the $z \to \infty$ limit. As we described around equation (\ref{eq:fixed}) above, this leads to a temperature dependence of the entropy density of $s \sim T^\eta$. Our backgrounds and equations of motion behave well under this limit. Specifically, the metric (\ref{eq:metric}) becomes
\be\label{eq:nicegeom}
ds^2 = \frac{L^2}{r^2} \left( - \frac{dt^2}{r^{4/\eta}} + g_0 \frac{dr^2}{r^2} + dx^2 + dy^2 \right) \,.
\ee
This spacetime is conformally related to $AdS_2 \times \R^2$, as we might have anticipated. The extent to which such spacetimes will share the properties of $AdS_2 \times \R^2$ will depend on the probe. Probes that are not sensitive to the conformal factor, such as conformally coupled fields, will see very similar physics. Probe fermions were considered in these backgrounds in \cite{Iizuka:2011hg}. Here we shall look at the transverse current spectral densities, that probe the entire charged sector. It was found in \cite{Gouteraux:2011ce} that these geometries can be uplifted to an Anti-de Sitter spacetime times a plane. This lift presumably underlies the simple correlation functions we will now exhibit.

The zero temperature entropy density of $AdS_2 \times \R^2$ also manifests itself as an extensive entanglement entropy \cite{Swingle:2009wc}. The geometries we are considering, with $\eta > 0$, do not have a zero temperature entropy density and also do not have an extensive entanglement entropy. The entanglement entropy of a region with perimeter $R$ obeys the `boundary law'
\be\label{eq:se}
S_E \sim \frac{R}{\e} \,,
\ee
with no extensive subleading term. Here $\e$ is a short distance spatial cutoff in the field theory. See for instance the formulae in \cite{Huijse:2011ef}.

Computing the entanglement entropy of a long strip, with lengths $R_y \gg R_x$, in these backgrounds leads to a curious property. A minimal surface connecting the lines at the boundary only exists for a specific separation, $R_x = R_\text{crit} = \sqrt{g_0} \pi/4$. If the locally critical scaling geometry is the IR limit of an asymptotically AdS spacetime, then in the full spacetime, a connected minimal surface only exists for separations $R_x \leq R_\text{crit.}$. As $R_x \to R_\text{crit.}$, the minimal surface droops increasingly farther into the IR, so that $r_\text{max.} \to \infty$. The minimal surface for separations $R_x > R_\text{crit.}$ is presumably disconnected, with two surfaces falling into the IR at constant separation. This is reminiscent of the behavior of holographic entanglement entropy in confined phases \cite{Nishioka:2009un}, and yet our phase is not confined: the entropy density $s \sim T^\eta$, and we will shortly exhibit spectral weight at low frequencies. Similar behavior was also found for the entanglement entropy of little string theory \cite{Ryu:2006ef}. Future work will need to clarify the physical interpretation of the lengthscale $R_\text{crit.}$ in these geometries. A related discussion for circular regions, which behave slightly differently, appears in \cite{Liu:2012ee}.

We now proceed to solve the perturbation equations describing transverse current fluctuations.
Similarly to before, we rescale the fields:
\be
\psi_1(r) =  \frac{1}{\sqrt{1+\eta}} \frac{1}{Z_0} \frac{\w^{\eta-1}}{k^{\eta}} \phi_1(r) \,, \qquad \psi_2(r) = \phi_2(r) \,,
\ee
and change the radial variable to
\be\label{eq:rr}
r = \left( \frac{k}{\w} \right)^{\eta/2} \rho^{\eta/2} \,.
\ee
Let us also rescale the momentum by
\be
\hat k = \frac{\sqrt{1 + \eta}}{V_0} k \,.
\ee
In the double scaling limit, the perturbation equations become
\be
\phi_1'' + \frac{\eta}{\r} \phi_1' + \frac{(1+\eta)}{\r^{\eta}\left(\r^2-1\right)} \phi_2' +\left(\hat k^2 \frac{(\r^2-1)}{\r^2} - \frac{2}{\r^2-1} \right) (1+\eta) \phi_1 = 0 \,,
\ee
and
\be
\phi_2'' + \frac{(1-\r^2)\eta-2}{\r (\r^2-1)}\phi_2'
- 2 \r^{\eta-2} \phi_1' + \frac{4 \r^{\eta-1}}{\r^2-1} \phi_1 + \hat k^2 \frac{(1+\eta) (\r^2-1))}{\r^2} \phi_2 = 0 \,.
\ee
We see that, in contrast to our rescaled equations at finite $z$, (\ref{eq:p1}) and (\ref{eq:p2}) above, the momentum appears on its own, not divided by powers of the frequency. Therefore, in these locally critical geometries we can take $\w \to 0$ at fixed $k$ without being pushed into a WKB regime.

While these equations are not identical to those describing perturbations about $AdS_2 \times \R^2$, the decoupling trick used by \cite{Kodama:2003kk,Edalati:2010hk} also works in this case. Thus we define
\be\label{eq:pm}
\phi_\pm = \frac{\r^{2-\eta/2}}{(\r^2-1)} \left( \r^{\eta-2} \left\{(\r^2-1)\left(-1 \mp \sqrt{1 + 2 \hat k^2} \right)  -2 \right\} \phi_1 + \phi_2' \right) \,.
\ee
These fields satisfy the decoupled equations
\be
\phi_\pm'' = \frac{1}{\r^2}\left(\pm (1+\eta) \sqrt{1 + 2 \hat k^2} +
1 + \frac{1}{4} \eta(2 + \eta) - (1+\eta) \hat k^2 (\r^2-1)
\right) \phi_\pm \,.
\ee
The decoupled equations are now easily solved in terms of Bessel functions. The solution satisfying infalling boundary conditions at the horizon is naturally expressed in terms of the Hankel function
\be\label{eq:han}
\phi_\pm = \sqrt{\r} H^{(1)}_{\nu_\pm} \left( \sqrt{1+\eta} \hat k \r \right) \,.
\ee
The indices 
\be\label{eq:ind}
\nu_\pm = \frac{1}{2} \sqrt{5 + 4 (1 + \eta) \hat k^2 + \eta (2+ \eta) \pm 4(1+\eta) \sqrt{1 + 2 \hat k^2}} \,.
\ee
When $\eta = 0$, these indices are the same as those found in \cite{Edalati:2010hk} for $AdS_2 \times \R^2$, after rescaling the momentum. The geometry (\ref{eq:nicegeom}) indeed becomes $AdS_2 \times \R^2$ in that case. For general $\eta$, the indices (\ref{eq:ind}) and metric (\ref{eq:nicegeom}) are inequivalent to the $AdS_2 \times \R^2$ values.

The computation of the retarded Green's functions and associated spectral density now follows very closely the $AdS_2 \times \R^2$ case \cite{Edalati:2010hk}, the main difference being the values of $\nu_\pm$. Expanding the infalling solution (\ref{eq:han}) near the boundary $\r \to 0$, and re-introducing the original radial coordinate $r$, see (\ref{eq:rr}), the retarded Green's functions for the two decoupled modes are read off by comparing with
\be
\phi_\pm \propto r^{1/\eta} \left(r^{-2 \nu_\pm/\eta} + {\mathcal{G}}^R_\pm(\w,k) r^{2 \nu_\pm/\eta} \right) \,, \qquad (\text{as $r \to 0$}) \,.
\ee
The retarded Green's functions are found to have real and imaginary parts scaling with the same power of $\w$, so that the spectral density is
\be\label{eq:zbig}
\text{Im} \, {\mathcal{G}}^R_\pm(\w,k) \; \propto \; {\mathcal{G}}^R_\pm(\w,k) \; \propto \; \w^{2 \nu_\pm(k)} \,.
\ee
This result describes the exact frequency dependence of the Green's function for the scaling geometry. The overall normalization includes an additional nonsingular momentum dependence \cite{Edalati:2010hk}. If the scaling geometry describes the IR of a full asymptotically AdS spacetime, for instance, then the above formula will describe the leading low frequency behavior of the spectral density in the full theory. The transverse current is dual to the bulk field $\psi_1$, which is a linear combination of $\phi_\pm$ according to (\ref{eq:pm}). At low frequencies, the more singular of the two exponents in (\ref{eq:zbig}) dominates and therefore we will have
\be
\text{Im} \, G^R_{J_\perp J_\perp}(\w,k) \; \propto \; \w^{2 \nu_-(k)} \,.
\ee

Taking a low momentum limit, $k \to 0$, we obtain $\nu_- \to \half |\eta-1|$. This shows that a nonzero $\eta$ can allow the locally critical scaling dimension of the transverse current to become smaller than in the case of $AdS_2 \times \R^2$, where $\nu_-$ is bounded below by $\half$. If a similar effect occurs for the charge density $J^t$, then this may lead to very strong umklapp and impurity scattering in these cases \cite{Hartnoll:2012rj}. Computing the charge density correlator is a more challenging computation, as more modes are coupled in the sound channel, but should not be substantially more difficult than the $AdS_2 \times \R^2$ case \cite{Edalati:2010pn}. It is also of interest to see if the finite temperature Green's functions can be computed explicitly. In the $AdS_2 \times \R^2$ case \cite{Hartnoll:2012rj}, the finite temperature Green's function is related to the zero temperature result by the underlying $SL(2,\R)$ symmetry \cite{Faulkner:2011tm}.

As anticipated, we see in (\ref{eq:zbig}) that taking $z=\infty$ leads to a spectral weight at low frequencies and finite momenta that is not exponentially suppressed, in contrast to what we found at finite $z$. The result is qualitatively similar to the case of $AdS_2 \times \R^2$, but with different exponents $\nu_\pm$ and with an entropy density that goes to zero with the temperature.

The cost of generalizing $AdS_2 \times \R^2$ in the way we have done is that, in common with most hyperscaling metrics, 
the spacetimes have divergent curvatures in the far IR as $r \to \infty$. As usual, this likely indicates that new physics will arise at the lowest energy scales.

\section{Discussion}

A dual field theoretic understanding of charged horizons as a finite density state of matter remains elusive. If the charge carriers are to be thought of as `bosonic', then one must understand why these bosons have not condensed. If the charge carriers are to be thought of as `fermionic', then one must understand where the finite momentum low energy spectral weight has gone. Perhaps these will turn out not to be the correct questions.

Low energy theories of critical bosons coupled to Fermi surfaces are naturally formulated in terms of `patches' on the Fermi surface, e.g. \cite{sungsik, Metlitski:2010pd, Metlitski:2010vm}. The patch picture is consistent insofar as there is a certain degree of decoupling between patches at the lowest energy scales. However, in currently existing finite density gravity duals this patch structure is not manifest. The emergent scaling symmetry is towards the origin of momentum space rather than towards a Fermi surface. The exception to this statement is if $z=\infty$, wherein all momenta are independently critical at low energies. Thus, of known gravity backgrounds, these locally critical theories come the closest to the patch structure of more conventional field theoretic approaches. This fact manifested itself in our computation in that these were the only cases in which the gravity dual had spectral weight at low energies and nonzero momenta.

Even in the locally quantum critical cases, one does not see the `back' of the Fermi surface, as all momenta are critical at low energy. It may be that seeing sharper Fermi surface-like physics necessarily requires quantum gravitational or string theoretic input. A recent interesting result in this direction is the momentum space structure seen by string probes in \cite{Polchinski:2012nh}. To make the most of the power of holography, it would be nice to understand how Fermi surface physics can be exhibited at leading order in the bulk classical gravity limit. One type of momentum space structure that has been found in the classical bulk limit are instabilities towards spatially modulated phases \cite{Nakamura:2009tf, Donos:2011bh}. Perhaps this is a signal that Fermi surfaces in theories with gravity duals are generically unstable towards forming charge density waves?

\section*{Acknowledgements}

It is a pleasure to acknowledge discussions with Blaise Gouteraux, Liza Huijse, Juan Jottar, Shamit Kachru, Subir Sachdev and Eva Silverstein. E.S. is supported by the Stanford Institute for Theoretical Physics under NSF Grant PHY-0756174. The work of S.A.H. is partially supported by a Sloan research fellowship.

\end{document}